\documentstyle[prl,aps,graphicx]{revtex}
\begin{document}
\draft

%%  \twocolumn
%%   \narrowtext

\title{Impact of Optical Modes on the Pairing Potential in Bilayer Cuprates}
\author{I.A. Larionov$^{*}$ and M.V. Eremin}
\address{Magnetic Radiospectroscopy Laboratory, Department of Physics,
Kazan State University, 420008 Kazan, Russia}
\maketitle
\date{\today}

\begin{abstract}

Superconducting transition temperature is calculated for differently doped
bilayer cuprates. Superexchange is assumed to be the dominating mechanism of
high-temperature superconductivity, but the contribution from the phonon
potential is not negligible, which qualitatively explains the observed weak
isotopic effect. The calculated value 2$\Delta _{max}$/$k_B T_C \simeq 4.5$
is close to the experiment in the case of optimum doping.

\end{abstract}

$$ $$

[*]  e-mail: {\it iL} @ ksu.ru

$ $

Journal ref:

$ $

Izvestiya Rossijskoj Academii Nauk, Physics Series, Vol.~{\bf 62}, No.~8,
pp.~1518-1521 (1998).

$ $

Corrected English translation: Bulletin of the Russian Academy of Sciences:
Physics, Vol.~{\bf 62}, No.~8, pp.~1228-1230 (1998).

$ $


%%  \twocolumn
%%   \narrowtext

\section{Introduction}

The High Temperature Superconductivity (HTSC) was discovered 11 years ago.
None of the proposed HTSC mechanisms has been commonly accepted yet (see
\cite{Loktev}). Unlike the case in normal low-temperature superconductors,
the energy gap in HTSC materials is different at different points of the
Brillouin zone. This dependence is important, for the understanding of the
HTSC mechanism.

In \cite{My_JETPL_1995,Physica_C_1997} we found solutions of the
Bardeen-Cooper-Schrieffer (BCS) equations taking copper spin superexchange
into consideration. This interaction was found to lead to a reasonable
critical temperature $T_C$ and $d$-symmetry of the energy gap in
superconducting cuprates. The conclusion about the $d$-symmetry of the order
parameter was confirmed by recent studies of photoemission spectra
\cite{Ding_ARPES_1996}. In the present work we additionally consider the
electron - phonon interaction with optical buckling $A_{1g}$, $B_{1g}$ and
breathing $A_{g}$, $E_{g}$ modes to explain the weak isotope effect in these
compounds.

\section{Potential of current carriers pairing through optical buckling
and breathing modes}

In the compound YBa$_{2}$Cu$_{3}$O$_{7-x}$ holes interact most strongly with
an electric field perpendicular to the CuO$_{2}$ plane. This field is mainly
induced by triply charged yttrium ions \cite{Devereaux}. The operator of
binding energy with oscillations perependicular to the CuO$_{2}$ plane has
the form

\begin{equation}
H_{h-ph}=e \sum_{n\sigma} {\Big \{ }  {\bf E}_x {\bf u}_x \left ( a{\bf n} +
\frac{a {\bf x}}{2} \right ) p^{\sigma \sigma}_{nx} + {\bf E}_y {\bf u}_y
\left ( a{\bf n}+\frac{a {\bf y}}{2}\right ) p^{\sigma \sigma}_{ny}{\Big\} },
\end{equation}
where $p^{\sigma \sigma}_{nx}$ and $p^{\sigma \sigma}_{ny}$ are the Hubbard
operators of oxygen holes, ${\bf u}_{x} (n) $ and ${\bf u}_{y} (n) $ are the
displacements vectors of O(2) and O(3) positions in a unit cell with the
number $n$, ${\bf x}$ and  ${\bf y}$ are unit vectors of the axes $a$ and
$b$ respectively, $a$ is the lattice constant, and $E_{x} = 1.2\cdot 10^8$~
V$\cdot$cm$^{-1}$, $E_{y} = 1.5\cdot 10^8$~V$\cdot$cm$^{-1}$ are the
components of electric field along the the axis $c$, calculated in
\cite{Eremin_Lavizina}. For simplicity, we put $E = E_{x} = E_{y} =
1.35\cdot 10^8$~V$\cdot$cm$^{-1}$. Then, calculating the commutator $\left [
\Psi_{\bf k}^{\downarrow, pd}, H_{h-ph} \right ]$ and using the expression

$$\left [ \Psi_{\bf k}^{\downarrow, pd}, H_{h-ph} \right ] = \sum_{\alpha,
{\bf q}} V^{\alpha} ({\bf q}) \Psi_{{\bf k}-{\bf q}}^{\downarrow, pd}
(b_{\bf q} + b_{-{\bf q}}^{+}) ,$$
where $\Psi_{\bf k}^{\downarrow, pd}$ is
the quasiparticle operator of singlet-correlated oxygen holes, $b_{-{\bf
q}}^{+}$ and $(b_{\bf q}$ are the phonon creation and annihilation
operators, one finds

\begin{equation}
V^{\alpha}({\bf q})= \frac {e}{2} E \sqrt{\frac {\hbar}{2m\omega_{\alpha}} }
\left [ \cos \left ( \frac {q_x a}{2} \right ) \pm \cos \left (\frac
{q_y a}{2} \right) \right] ,
\end{equation}
where $\omega_{\alpha} = \omega_{{\mathrm A}_{1g}} = $440~cm$^{-1}$ and
$\omega_{{\mathrm B}_{1g}} = $340~cm$^{-1}$ \cite{Thomsen_Cardona}. The plus
and minus signs correspond to the modes A$_{1g}$ and B$_{1g}$ respectively,
$e$ is the electron charge, and $m$ is the mass of the unit cell. Since the
difference between $\omega_{{\mathrm A}_{1g}}$ and $\omega_{{\mathrm
B}_{1g}}$ is not so important, they are put equal to $\omega_{{\mathrm G}}
= $400~cm$^{-1}$.

Using the Frohlich procedure, the potential of current carriers interating
through the phonon field of buckling modes A$_{1g}$ and B$_{1g}$ is written
as
\begin{equation}
G({\bf k}^{\prime} -{\bf k}) = 2G_0^2 \frac { 1+\frac {1}{2} \left[ \cos
\left(k_x^{\prime} -k_x\right)a + \cos \left(k_y^{\prime} -k_y \right)a
\right] } {\left( \varepsilon_{\bf k^{\prime}} - \varepsilon_{\bf k} \right)^2
- \left(\hbar \omega_{G} \right)^2 } \hbar \omega_{G},
\end{equation}
where $G_0 =$~35~meV.

Assuming that the constant of interaction with the modes A$_g$ and E$_g$
(and their frequencies) are identical, the contribution to the pairing
potential is written as
\begin{equation}
B({\bf k}^{\prime} -{\bf k}) = 2B_0^2 \frac { 1- \frac {1}{2} \left[ \cos
\left(k_x^{\prime} -k_x\right)a + \cos \left(k_y^{\prime} -k_y \right)a
\right] } {\left( \varepsilon_{\bf k^{\prime}} -\varepsilon_{\bf k} \right)^2
- \left(\hbar \omega_{B} \right)^2 } \hbar \omega_{B}.
\end{equation}
According to the estimate of \cite{Eremin_Naturforsch}, the constant $B_0$
is about 60~meV. The frequencies $\omega_{{\mathrm A}_{g}}$ and
$\omega_{{\mathrm E}_{g}}$, taken from \cite{Litvinchuk}, are about
480~cm$^{-1}$.

\section{Equation for the gap}

The BCS equation for the energy gap, taking into account the interaction via
the phonon field and the superexchange between copper spins, has the form
\begin{equation}
\Delta_{{\bf k}^{\prime}} = \sum_{\bf k} {\Big \{ } P^2 \left [
G({\bf k}^{\prime} -{\bf k}) + B({\bf k}^{\prime} -{\bf k})\right ]
\theta(\hbar \omega_{D}-|\varepsilon_{\bf k^{\prime}} -\varepsilon_{\bf k}|)
- J^{dd}({\bf k}^{\prime} -{\bf k}) {\Big \} }
\left < \Psi_{\bf k}^{\downarrow,pd} \Psi_{-{\bf k}}^{\uparrow,pd} \right >,
\end{equation}
where $J^{dd}({\bf q})$ is the Fourier transform of the superexchange
parameter $J = 57 $~meV, $$ J^{dd}({\bf q}) =2J \left ( \cos q_x a +\cos q_y
a \right ),$$ $\omega_D = 500$~cm$^{-1}$ is the Debye frequency, and
$P=\frac {1}{2}(1+x)$. The correlation function is given by

$$ \left < \Psi_{\bf k}^{\downarrow,pd} \Psi_{-{\bf k}}^{\uparrow,pd} \right
> = -\frac {\Delta_{\bf k}}{2E_{\bf k}} \tanh \left (\frac{E_{\bf k}}
{2 k_B T} \right ),$$  where $E_{\bf k} = \sqrt{(\varepsilon_{\bf k}-\mu)^2
+ \Delta_{\bf k}^2}$. The chemical potential $\mu$ and the hole
concentration $x$ for two copper sites in the bilayer (assuming that the
antibonding band is ampty) are related as
$$ x = P \sum_{\bf k} \left [ \exp \left ( \frac {\varepsilon_{\bf k}
-\mu}{k_B T} \right ) +1 \right ] ^{-1}.$$
At $x=0.33$ the chemical potential is placed on 10~meV below the saddle
singularity peak in the density of states. The dispersion $\varepsilon_{\bf
k}$ is chosen as

\begin{equation}
\varepsilon_{\bf k} =P\left [2t_1 \left ( \cos k_x a + \cos k_y a \right )
+4t_2 \cos k_x a \cos k_y a +2t_3 \left ( \cos 2k_x a +\cos 2k_y a \right )
 \right ],
\end{equation}
where  $t_1$, $t_2$, and $t_3$ are the effective hopping integrals. All the
calculations are carried out at $t_1 =70$~meV, $t_2 =0$, and $t_3 = 5$~meV.

  \twocolumn
   \narrowtext

\begin{figure}  [tbp]
  \centering \includegraphics[width=0.99\linewidth] {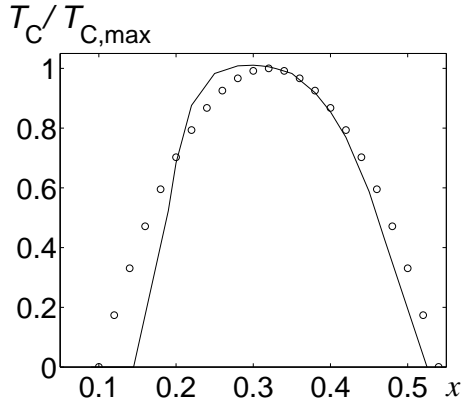}
\caption{Temperature dependence of the superconducting transition
temperature in YBa$_2$Cu$_3$O$_{7-y}$ with doping: solid line (calculations)
and the experimental points are sketched by open circles.}
\label{NS_Fig}
\end{figure}  %%%   Figure 1

Figure 1 shows the dependence of the superconducting transition temperature
of YBa$_2$Cu$_3$O$_{7-y}$ with doping. It is evident that the calculations
qualitatively conform with the experimental data (normalized by the
expression $T_C/T_{C,max} = 1-82.6(x-0.32)^2$ \cite{Williams_Tallon}).
However, their noncoincidence remains unexplained. Numerical solution of the
energy gap equation yields a $d$-type symmetry of the order parameter
$\Delta_{\bf k} = \Delta_0 ( \cos k_x a - \cos k_y a)$ and a nonstandard
value of $2\Delta_{max}/k_B T_C \approx 4.5$ with the superconducting
transition temperature $T_{C,max} \sim 100$~K. Note that solutions of (5)
with $J=0$ lead to the $s$-type symmetry of the gap, $\Delta_{\bf k} =
\Delta_0 ( \cos k_x a + \cos k_y a)$. The conclusion of
\cite{Nazarenko_Dagotto} that the $d$-symmetry of the gap arises from
pairing only through optical buckling modes seems to be groundless.

\section{Conclusion}

Self-consistent solutions of the BCS gap equation are found in the class of
short-range pairing potentials for various chemical potentials. When the
Fermi level $\varepsilon_{F}$ is near the bottom(top) of the band, the
solutions correspond to the $s$-type pairing, while for $\varepsilon_{F}$ in
the center of the band the solution are related to the $d$-type. The
short-range potentials considered are (i) superexchange interaction, (ii)
interaction of current carriers via optical oscillations, and (iii)
breathing and buckling modes of oxygen atoms in CuO$_2$ planes. The
superconducting transition temperature is calculated for various oxygen
indices $x$.

The dominant HTSC mechanism is a superexchange. However, the contribution of
the phonon pairing is not negligible, that qualitatively explains the
observed weak isotope effect. The calculated value $2\Delta_{max}/k_B T_C
\approx 4.5$ is close to the experiment.

The work was partially supported by the Federal Program "Superconductivity"
(Grant 94029) and the Russian Foundation for Basic Research under Project
Code 97-02-16235.

\end{document}